# A Functional Programming Language with Versions

Yudai Tanabe[a], Luthfan Anshar Lubis[a], Tomoyuki Aotani[b], and Hidehiko Masuhara[a]

a Tokyo Institute of Technology, Tokyo, Japan

b Mamezou Holdings Co., LTD., Tokyo, Japan

**Abstract** While modern software development heavily uses versioned packages, programming languages rarely support the concept of versions in their semantics, which makes software updates more bulky and unsafe. This paper proposes a programming language that intrinsically supports versions. The main goals are to design core language features to support multiple versions in one program and establish a proper notion of type safety with those features. The proposed core calculus, called Lambda VL, has versioned values, each containing different values under different versions. We show the construction of the type system as an extension of coeffect calculus by mapping versions to computational resources. The type system guarantees the existence of a valid combination of versions for a program. The calculus enables programming languages to use multiple versions of a package within a program. It will serve as a basis for designing advanced language features like module systems and semantic versioning.

# 1 Introduction

It is common to use versioned packages in software development. A *package* is a unit of software components developed and managed by the software developers themselves or by external developers. A *version* is usually a chain of numbers that distinguishes an implementation of an evolving package. A newer version usually improves the older ones by adding features, improving performance, or fixing bugs. Developers can notice the availability of newer implementations of externally developed packages by checking their version numbers and decide whether they should replace the packages currently in use with the new ones.

A new version of a package can either be *compatible* or *incompatible* (or with "breaking changes") from older versions. When it is compatible, we can replace an older version with the new one without a problem. Otherwise, we would need to modify our program in order to use the newer version [13, 28].

## 1.1 Existing Techniques to Use Multiple Versions of a Package

A new yet incompatible version of a package is an ambivalent thing: while it brings benefits, it comes with a certain amount of cost of modifying the programs that are using the package. The cost can be huge when a program happened to require a new and an older version simultaneously [3]. Though many systems, programming languages, and execution environments allow programs to use only one version for each package simultaneously, there are many techniques to relax this restriction for more flexibility. We can classify those techniques in terms of the unit in which such a restriction is posed.

- *Device*: For packages that are only allowed to exist in one version on a device (e.g., operating system standard libraries like the C standard library (libc)), OS-level virtualization software enables one to use different versions in the virtualized environment. For example, traditional UNIX-like operating systems only allow one version of the OS standard library; OS-level virtualization software such as Docker [20] and QEMU [4] can provide environments with different versions.
- *Process*: For a package (or a library) linked to a program, a dynamic loading mechanism can provide a different version for a different process of the program. For example, with the shared library mechanism in UNIX-like operating systems, a compiled program can run with a different version of a library by providing a different load path. There are also more advanced mechanisms, e.g., OSGi [1], that automatically load appropriate versions.
- *Package*: For a package required by two modules of a program, some package managers allow to load two different versions of the required package. Such features can be found in npm[1] for JavaScript, cargo[2] for Rust, and Maven[3] with the shade plugin[4] for Java and are described in section 2.3.

By following this technology trend that supports multiple versions with a finer computation unit, we investigate programming languages that support multiple versions inside a module. In contrast to previous technical efforts that focused on avoiding the simultaneous use of multiple versions, we focus on programming to simultaneously use multiple versions of a package. In our proposed language, the versions are held in units of expressions so that each expression can call a different version of the implementation. We establish a foundation for more freely combining and controlling different versions through a language-based approach.

As there are few attempts to develop such languages, it is not obvious what language abstractions are suitable to represent multiple versions and what kind of safety we can guarantee. As a first step, we propose a calculus called $\lambda_{\mathrm{VL}}$, which models core language features for such programming languages.

### 1.2 Versions within a Language Semantics

$\lambda_{\mathrm{VL}}$ supports multiple versions at the value level and has a type system that guarantees a program to use values consistently with respect to versions. Our ideas are (1) introducing *versioned values*: records of multiple values distinguished by their version, and (2) statically checking whether versions of functions and arguments agree through a type system.

(1) *Versioned values* represent multiple versions of a computation and bundle them as a single value. Versioned values allow us to abstract multiple versions of values. For example, the versioned value $\{v_1 = \lambda x.x, v_2 = \lambda x.x + 1\}$ represents a versioned function whose initial implementation ($v_1$) is an identity function and its next implementation ($v_2$) is a successor function.

Applying a versioned function to a versioned value results in a versioned value. This versioned value consists of version-specific terms obtained by applying version-specific functions to the corresponding values in the versioned value, in a version-wise manner. For example, we obtain $\{v_1 = 1, v_2 = 3\}$ if we apply $\{v_1 = \lambda x.x, v_2 = \lambda x.x + 1\}$ to $\{v_1 = 1, v_2 = 2\}$. If a function and its arguments have a different set of versions, the application is calculated on the common part of each version set. For example, we obtain $\{v_1 = 1\}$ if we apply $\{v_1 = \lambda x.x, v_2 = \lambda x.x + 1\}$ to $\{v_1 = 1\}$.

(2) To guarantee type safety, we develop a calculus called $\lambda_{\mathrm{VL}}$ based on the *coeffect calculus* [7]. A coeffect calculus is a type system derived from linear type systems [17, 33] and a type system scheme for analyzing the usage of various computational resources, not just the number of times a variable is used. Just as other coeffect calculi track their computational resources, $\lambda_{\mathrm{VL}}$ attaches version numbers to types, such as $x : \Box_{\{v1,v2\}} T$, meaning $x$ is a variable of type $T$ and computable under versions $v_1$ and $v_2$. The $\lambda_{\mathrm{VL}}$ type system collects the annotated type information for the program and calculates the set of versions needed to run the program. The type system ensures that there is at least one consistent version where the given program can be evaluated.

[1] https://www.npmjs.com/ (February 1, 2021)
[2] https://doc.rust-lang.org/cargo/ (February 1, 2021)
[3] https://maven.apache.org/ (February 1, 2021)
[4] https://maven.apache.org/plugins/maven-shade-plugin/ (February 1, 2021)

**Table 1** Structural change: availability of functions in GDK 3

| version | gdk_screen_get_n_monitors | gdk_display_get_n_monitors |
|---|---|---|
| $< 3.22$ | available | not available |
| $\geq 3.22$ | deprecated | available |

**Table 2** Behavioral change: reliability of an alarm time in Android API

| version | set | setExact |
|---|---|---|
| $< 19$ | exact | not available |
| $\geq 19$ | inexact | exact |

In summary, we make the following contributions:

- We introduce versions into a programming language and demonstrate that the concept alleviates the dependency-hell problem (section 2, section 3).
- We develop a type system of $\lambda_{\mathrm{VL}}$ (section 4).
- We propose the notion of type safety of programs with multiple versions and prove it (section 5).

In the non-technical sections of this paper, we discuss related research in section 6 and further work in section 7.

## 2 Incompatibility Problems

We focus on incompatibilities and the existing methods to manage them. For projects that depend on frequently updated open-source packages, each package is continually evolving, hence compatibility issues are inevitable [2, 3, 5, 13, 14, 28].

### 2.1 Types of Incompatibility

Code changes can cause the following two incompatibilities:

- Structural incompatibilities:
  Multiple versions of a package provide different sets of definitions, such as function names and data structures.
- Behavioral incompatibilities:
  Multiple versions of a package provide the same set of definitions, but their behaviors are different.

#### 2.1.1 Structural Incompatibilities

Structural incompatibilities are caused by the addition and removal of definitions, internal changes to data structures, renaming, etc. Table 1 shows an example of a structural incompatibility in GIMP Drawing Kit (GDK). GDK is a C library for creating graphical user interfaces and is used by many projects, including GNOME.

If we suppose that deprecated functions are unavailable, version 3.22 is structurally incompatible with version 3.20 because the former lacks `gdk_screen_get_n_monitors` that is available in the latter. GDK versions prior to 3.22 provide `gdk_screen_get_n_monitors` that tells the number of physical monitors connected. However, versions 3.22 and later provide the same functionality function `gdk_display_get_n_monitors` and deprecate `gdk_screen_get_n_monitors`. When we upgrade GDK to version 3.22 and build software that uses this function without modifying anything, the build system will give us an undefined reference error. With a static type check, the programmer will be informed of the incompatibility problem as a compilation error.

#### 2.1.2 Behavioral Incompatibilities

Code changes may also cause behavioral incompatibilities that include the additions, removals, and changes of side effects, even if there is no change in name or type. Table 2 shows an example of the behavioral incompatibility in the Android Platform API (henceforth Android API). The Android API is the standard library mostly written in Java, and its version is synchronized with that of the Android OS.

Prior to version 19,[5] the Android API provided the set method in the `AlarmManager` class that schedules an alarm at a specified time. However, since version 19, the Android API has changed its behavior for the sake of power management. Despite having the exact same name and type definitions, set no longer guarantees accurate alarm delivery. For developers who require accurate delivery, the method `setExact` is provided instead.

### 2.2 Situations Where Incompatibility Problems Occur

Incompatibility problems are more likely to occur when there are transitive dependencies among subcomponents [2, 5]. Since it is common to develop with multiple packages, programmers often encounter this kind of problem [12]. Figure 1 shows a situation where the developer updates the software called *App*. The upper and the lower halves show the configurations before and after the update, respectively. On the upper half, *App* depends on Package *A* version 1 and Package *B* version 1, and also, Package *A* version 1 depends on Package *B* version 1. In this configuration, there is no incompatibility problem.

Suppose that the developer of *App* decided to switch the version of Package *A* from 1 to 2. Package *A* version 1 and 2 are compatible, but the dependency on Package *B* has been changed from version 1 to 2. Suppose that Package *B* version 1 and 2 are incompatible. Since *App* itself requires Package *B* version 1, it is impossible to use Package *A* version 2 from *App* without modifying *App* to use Package *B* version 2.

[5] The Android API uses *levels* instead of versions as identifiers for API revisions, but we will call them versions for consistency.

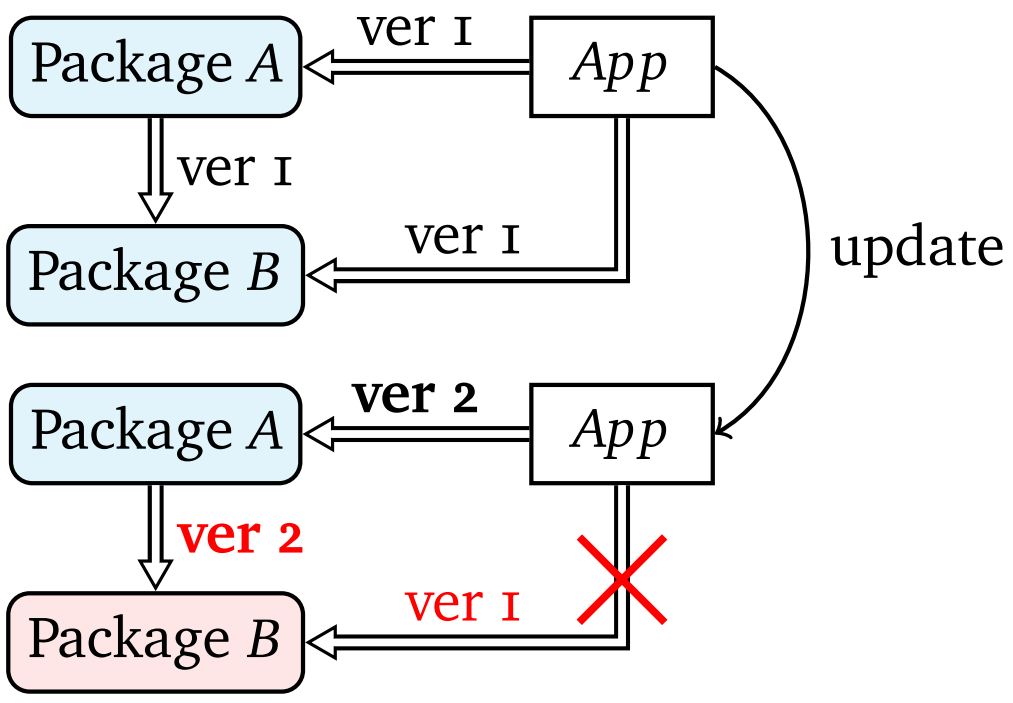

After the update, the two dependencies on Package *B* cause a conflict.

**Figure 1** Minimal configuration before (top) and after (bottom) the update that causes dependency hell.

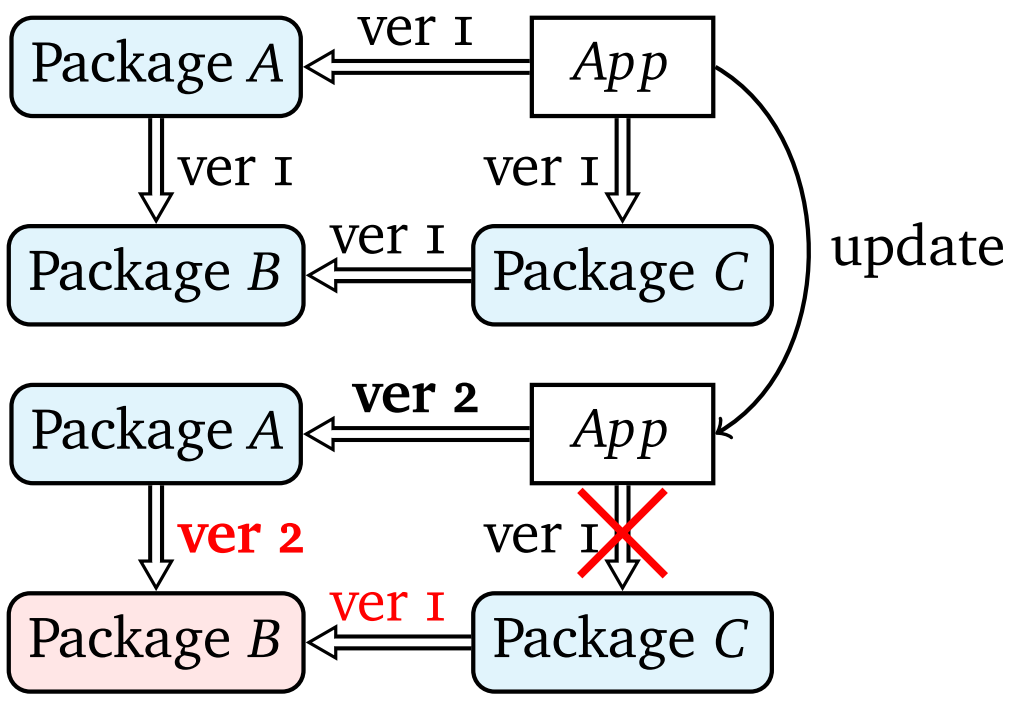

After the update, the two dependencies on Package *B* cause a conflict.

**Figure 2** Package configuration before (top) and after (top) the update cause dependency hell, where only transitive dependencies on Package *B*.

Even worse, if a program indirectly depends on two incompatible versions of a package, there is no way to fix the problem until the developers of intermediate packages catch up. As shown in fig. 2, if App used Package *C*, which in turn depends on Package *B* version 1, the *App* developer would have no way of resolving the conflict on their own.

### 2.3 External Tool Support for Dependency Conflict in Transitive Dependency

The external tool support for transitive dependency is based on the following ideas:

- Clarifying the dependencies using the notation of *semantic versioning* (SemVer) [27].
- Generating two separate copies of a dependent package when multiple packages have a common dependent package with SemVer-incompatible versions.

The semantic versioning recommendation [27] specifies that version numbers are basically a sequence that consists of major, minor, and patch versions separated by dots as in MAJOR.MINOR.PATCH. The semantic versioning strategy can explain to users of a package what types of changes will occur in the new release, and users can use version numbers as a guide to decide whether to accept the new release. For example, when incrementing the minor version of a package (e.g., 1.2.1 ⇒ 1.3.0), all changes should be backward compatible with previous versions. In contrast, incompatible code changes are permitted only when the MAJOR level is incremented (e.g., 1.2.1 ⇒ 2.0.0).

#### 2.3.1 Declarative Dependency Requirements

Some package managers, such as npm and cargo, use semantic versioning so that package users can automatically update dependent packages while maintaining compatibility. For this purpose, all package developers need to properly define the version requirements of their dependent package, called *dependencies*.

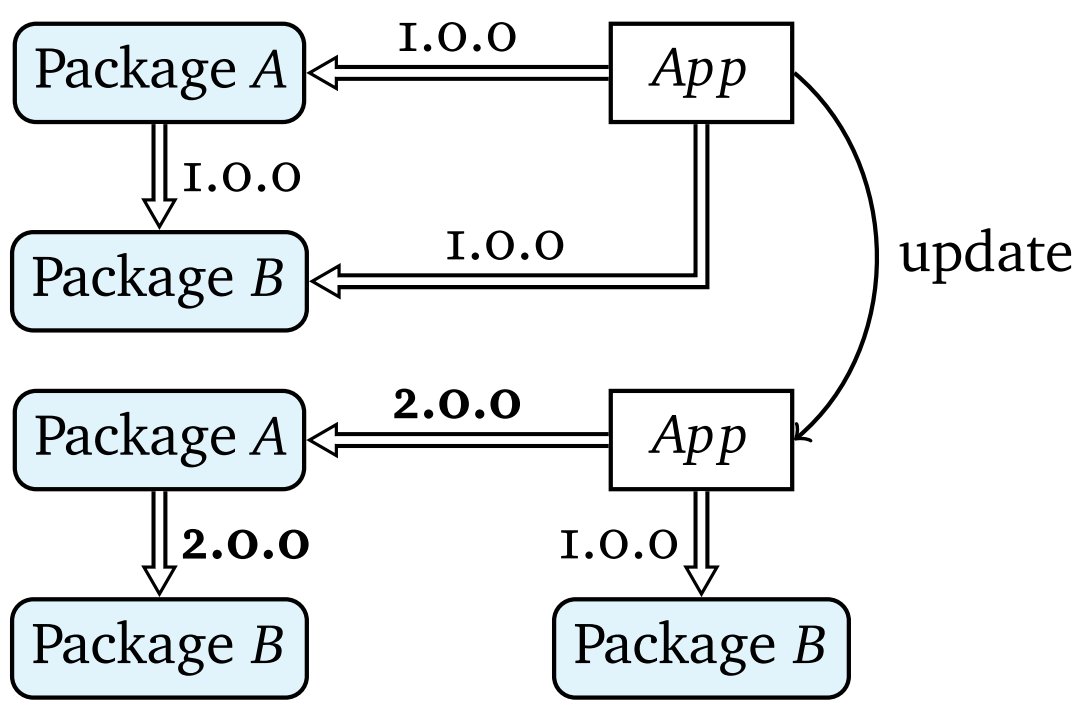

**Figure 3** Minimal package configuration in cargo to accept multiple versions of Package *B*.

Usually, dependencies can be described in the manifest file with each package using the SemVer notation. For example, similar to fig. 1, fig. 3 shows a situation where the developer updates software called *App*. The upper half of fig. 3 shows the same package configuration as the upper half of fig. 1. The cargo user can specify the dependencies of fig. 3 as follows:

```
[package]
name = "App"
[dependencies]
A = "1.0.0"
B = "1.0.0"
```

```
[package]
name = "A"
version = "1.0.0"
[dependencies]
B = "1.0.0"
```

In cargo, 1.0.0 means >=1.0.0 && <2.0.0. This requirement shows the range of versions that are backward compatible with 1.0.0 based on SemVer. Cargo collects these two manifest files and automatically retrieves the latest version that meets all requirements for each package. Assuming that all packages have only 1.0.0 and 2.0.0 for simplicity, the requirements above are satisfied by getting 1.0.0 for all packages.

#### 2.3.2 Name Mangling in External Tools

Modern, sophisticated package managers have a mechanism to alleviate the one-version-at-a-time policy by mangling/shading package names. As with fig. 1, when an *App* developer decides to switch Package *A* from version 1 to version 2 that requires Package *B* version 2, then the manifest file is modified as on the left below. Due to the modification, cargo refers to the manifest file of Package *A* version 2 as on the right.

```
[package]
name = "App"
[dependencies]
A = "2.0.0" // modified
B = "1.0.0"
```

```
[package]
name = "A"
version = "2.0.0" // updated
[dependencies]
B = "2.0.0" // dependency updated
```

In this example, *App* and Package *A* have a common dependency to Package *B*, but *App* and Package *A* require version >=1.0.0 && <2.0.0 and >=2.0.0 && <3.0.0 of Package *B*, respectively.

As shown in the lower half of fig. 3, cargo alleviates this by making two copies of Package *B*. The Rust compiler assigns a unique mangle symbol name to each program-ming unit using the package name and version. In this way, even if the same function name exists in different versions of a package, it is possible to determine the correct version of the function needed for each package. This solution merely replicates Package *B* into two completely different packages. However, it allows different versions of the same package to coexist in a dependency graph, in a sense relaxing the limitations of the one-version-at-a-time policy.

### 2.4 Problems Caused by Name Mangling

Name mangling is one reasonable mitigation measure, but it leads to new type-level incompatibilities [11, 32]. This approach collapses when values derived from different versions of a package are inevitably mixed in the same code. For example, consider the situation where Package *B* is a framework package that provides the Key type representing a cryptographic hash key, and Package *A* is a library that includes Key in its API. Suppose that the definitions of Key are the same in v1.0.0 and v2.0.0, and there are some breaking changes in other parts of the API as:

```
// package B v1.0.0
pub type Key = /* complicated */
```

```
// package B v2.0.0
pub type Key = /* Same as v1.0.0 */
```

Here, we simplify the code syntax, including function definitions, extern, semicolons, etc., as we do not want to address Rust-specific issues. By using the Key defined in Package *B*, the API in Package *A* and *App* are written as:

```
// package A v2.0.0
fn gen_key() -> Key // from v2.0.0
```

```
// App
let x : A::Key = gen_key() // type error!
```

Package *A* provides the gen_key function for generating hash keys. Note that the return type of gen_key is internally Key from v2.0.0.

In this example, the *App* program above will be rejected by the Rust type system because the Rust compiler gives v1.0.0 and v2.0.0 of the Key type completely different identifiers. The expected version for the Key structure is v1.0.0, but Package A v2.0.0 relies on Package B v2.0.0, so the gen_key function only returns the Key object in v2.0.0. Therefore, the programmer is eventually notified that the Key of v1.0.0 is not equal to the Key of v2.0.0 even if the definitions provided by both versions of Package *B* do not actually change at all.

The nature of this problem is that all interoperators must use compatible versions of the package. This fact has a tremendous impact on the ecosystem when Package *B* is a very widely used package such as a wrapper of libc and openssl[32]. Every time there is an update that does not guarantee backward compatibility, the developers of a very large number of packages need to collaborate to update their dependencies. For software with many dependencies, it is an ordeal to resolve dependency conflicts. Many developers are reluctant to upgrade dependencies unless they include a bug fix or other significant updates [3].

# 3 Approach: Programming with Versions

We are developping a programming language with a notion of versions to solve the problem discussed in section 2.4. We call a hypothetical surface language based on our ideas as *VL* here. A *VL* program can depend on multiple versions of a package, and static checks ensure that each value is dispatched to the appropriate implementation. We develop $\lambda_{\text{VL}}$ to be used as a core calculus for such surface languages. $\lambda_{\text{VL}}$ has the terms for combining multiple versions of definitions and the type system for identifying version-safe programs in multiple versions.

This section first describes the advantages of a language-based approach in section 3.1 and then illustrate core features through programming with $\lambda_{\text{VL}}$ in section 3.2.

## 3.1 Advantages of Language-Based Approach

The language-based approach has advantages over the package-based versioning approach as follows:

1. An appropriate version of a dependent implementation is selected for each expression.
   This feature allows programmers to write programs with a mixture of values from multiple versions, such as the program in section 2.4. Since package-based versioning requires a single dependent version to be defined outside of the language semantics, it is impossible to write such programs in a naive way.
2. Exhaustive static analysis can be performed on multiple combinations of package versions.
   In package-based versioning, the package version is determined by the language external to the language, so only programs with pre-fixed versions can be processed by the language system.

We will use the same program as in section 2.4 to illustrate the advantages. For the sake of discussion, suppose that VL is designed to bridge the usual programming style and our proposal as follows:

- Versions are held in the units of modules instead of packages.
- Module interfaces are written in a version-crossing manner and contains the information about which version each symbol is available.
- The interfaces are generated by the *VL* language system by aggregating information from each version of the module interface.

Based on the ideas above, we can rewrite the codes in section 2.4 in *VL* as follows:

```
// Package B v1.0.0 & v2.0.0
pub type Key = {Bv1: ... , Bv2: ... }
```

```
// Package A
fn gen_key -> Key = {Bv2: ... }
```

The interface of each package is rewritten in a version-crossing manner, where individual symbols are given version information, called *labels*. These are denoted as *Bv1* and *Bv2* and indicates the version in which the definition exists.

Note that we have given a notation for abstracting the two versions of Package *B*. This is totally different from the name mangling approach shown in figure 3, where the two versions of Package *B* are considered as completely different packages. In *VL*, the user of Package *B* can access to the both versions of definitions with the identifier *Key*. Which version of the computation is actually evaluated will be determined later by the type system as the available versions.

```
// App codes
let x : A::Key = gen_key() // v1.0.0 as a consistent version
```

Consequently, the *App* will be interpreted in *VL* as follows. The type system calculates the version shared by all the values in the data flow by using type checking and rejects them if they do not exist. In this example, the *Key* provided in Package *B* has definitions for both versions 1 and 2, and the gen_key provided in Package *A* has a definition for the *Key* in version 2 of Package *B*. As a result, it is apparent that the *App* code is available in the combination of version 2 of Package *A* and version 2 of Package *B*.

In this example, there was only one version combination available, but the type system computes all available version combinations. In this way, the language-based approach verifies programs in multiple version combinations.

### 3.2 Programming with Multiple Versions Mixed Together

As explained through the example, a language-based approach requires a version-crossing interface and a mechanism to know in which version each symbol is available. Therefore, it is important that the core language has the following two features.

1. A notation that expresses a value that consists of sub-values of multiple versions
2. A mechanism for analyzing which versions of a program may be available in more than one version

In this section, we will demonstrate how these features are achieved through the introduction of the core features of $\lambda_{\text{VL}}$.

#### 3.2.1 Versioned Values

$\lambda_{\text{VL}}$ is an extension of the coeffect calculus with versioned values that have multiple components tagged with versions. One way to construct versioned values is through *versioned records* $\{\overline{l_i = t_i}\}$[6],[7] We denote *labels* $(l_i)$ to distinguish the different versions of values and the values inside the versioned record are called *version-specific terms*. Versioned record provides a mechanism to write programs that are independent of a specific version. For example, to denote a default key length parameter that is 1024 and 4096 in versions 1 and 2 respectively, we can write

$$\{l_1 = 1024, l_2 = 4096\}.$$

[6] We will sometimes abbreviate a sequence as $\overline{*}$, i.e. $\overline{l_i}$ denotes $l_1, ..., l_n$ and $\overline{l_i = t_i}$ denotes $l_1 = t_1, ..., l_n = t_n$.

[7] Although our system, which we will describe in detail in section 4, explicitly states the default label such as $\{\overline{l = t} \mid l_k\}$, we omit it here for simplicity.

Another way to construct a versioned value is through *suspensions* $[t]$. The suspension $[t]$ promotes the term $t$ to a versioned value such as $[1]$ and $[\lambda x.x]$. The two constructors for versioned values delay the inside computation until a specific version is later determined.

To conduct a suspended computation, programmers can use *extractions* $t.l$. The extraction $t.l$ extracts the version-specific term according to the label $l$ from the versioned value returned by $t$. For example, consider the case where version 1 generates a key with a bit length of 1024, but version 2 generates one of length 4096. To generate a key with the appropriate bit length for each version and then retrieve the version-specific term in version 2, we can write

$$\{l_1 = \text{gen_key } 1024, l_2 = \text{gen_key } 4096\}.l_2.$$

#### 3.2.2 Versioned Function Application

Functions with different version-specific values across different versions are also represented as versioned values, called *versioned functions*. For example, as we saw in table 1 $v3.20$ and $v3.22$ of GDK 3 provide different named functions that have the same functionality. To define a new function that can retrieve the number of connected monitors in both versions, we can write

$$\{l_1 = \text{gdk_screen_get_n_monitors}, \\ l_2 = \text{gdk_display_get_n_monitors}\}.$$

Hereafter, we call this versioned function get_n_monitors.

To apply a versioned function, we need to pass a versioned value for the argument. Here, we can use *contextual let-binding* $\textbf{let } [x] = t_1 \textbf{ in } t_2$ to apply a versioned function. For example, we apply the versioned function get_n_monitors to the versioned value $\{l_1 = ()\}$ as follows:

$$\textbf{let } [f] = \text{get_n_monitors } \textbf{in let } [x] = \{l_1 = ()\} \textbf{ in } [f\ x] \tag{1}$$

This program first extracts the function gdk_screen_get_n_monitors from get_n_mo nitors and binds it to $f$; then it extracts the value () from $\{l_1 = ()\}$ and binds it to $x$.

#### 3.2.3 Versioned-Independent Programs

In the previous example, $\{l_1 = ()\}$ was bound to $x$ and had only one definition with $l_1$. However, if both the versioned function and versioned value have multiple definitions, the functional application should also be evaluated in multi-version contexts. We achieve this by using the suspension $[t]$. For example, we apply get_n_monitors to $\{l_1 = (), l_2 = ()\}$, both of which have two definitions with $l_1$ and $l_2$ as follows:

$$\textbf{let } [f] = \text{get_n_monitors } \textbf{in let } [x] = \{l_1 = (), l_2 = ()\} \textbf{ in } [f\ x]$$

This program returns a suspended computation that can return an integer value available in $l_1$ and $l_2$. Since the two version-specific terms in $\{l_1 = (), l_2 = ()\}$ are exactly the same, we can rewrite the above program as

$$\textbf{let } [f] = \text{get_n_monitors } \textbf{in let } [x] = [()] \textbf{ in } [f\ x].$$

#### 3.2.4 Types of Versioned Values

The type of a versioned value is denoted as $\Box_r T$. The index $r$, called the *version resources*, indicates which version-specific terms are available in the versioned value. This notion of type comes from coeffect calculus, and the exact method of calculating $r$ is based on the version resource semiring described later in section 4.

For example, assuming that both version-specific functions in get_n_monitors have type $\mathsf{Unit} \to \mathsf{Int}$, the above example programs are typed as follows:

$$\{l_1 = (), l_2 = ()\} : \Box_{\{l_1,l_2\}}\mathsf{Unit}$$
$$\text{get_n_monitors} : \Box_{\{l_1,l_2\}}(\mathsf{Unit} \to \mathsf{Int})$$

The type $\Box_{\{l_1,l_2\}}\mathsf{Unit}$ denotes that this versioned value has version-specific terms of type $\mathsf{Unit}$ and they are available in both versions $l_1$ and $l_2$.

The contextual let-binding $\textbf{let}\ [x] = t_1\ \textbf{in}\ t_2$ propagates the version requirements through the captured variable $x$. For example, the program of Eq. 1 is typed as follows:

$$\textbf{let}\ [f] = \text{get_n_monitors}\ \textbf{in}\ \textbf{let}\ [x] = \{l_1 = ()\}\ \textbf{in}\ [f\ x] : \Box_{\{l_1\}}\mathsf{Int} \tag{2}$$

where the result type has resource $l_1$ because get_n_monitors and $\{l_1 = ()\}$ only have $l_1$ as their shared labels.

Note that the extraction $t.l$ makes the type of $t$ lose the version resource. Once extracted, a version-specific term can be used with terms from other versions. For example, the type of the following program no longer have a version resource.

$$\textbf{let}\ [f] = \text{get_n_monitors}\ \textbf{in}\ \textbf{let}\ [x] = \{l_1 = ()\}\ \textbf{in}\ [f\ x].l_1 : \mathsf{Int} \tag{3}$$

#### 3.2.5 Ensuring a Consistent Version of the Computation

The $\lambda_{\text{VL}}$ type system ensures that all necessary versions of the implementation exist. In other words, if a program extracts a specific version of a value even though all the version values in the program do not have a shared label, the type system will reject such a program. The first example is a variant of Eq. 2.

$$\textbf{let}\ [f] = \text{get_n_monitors}\ \textbf{in}\ \textbf{let}\ [x] = \{l_3 = ()\}\ \textbf{in}\ [f\ x] : \Box_{\emptyset}\mathsf{Int}$$

The type system keeps track of the available versions of each variable by a set of labels in the context. In this example, it records $\{l_1, l_2\}$ for $f$, and $\{l_3\}$ for $x$. For each promotion, the type system calculates the shared version resource in the context that should be multiplied by the term. In the program type above, $[f\ x]$ is given version resource $\emptyset = \{l_1, l_2\} \cap \{l_3\}$, which indicates that there is no shared version available. It means no longer possible to extract any version-specific computation from this program and such extractions will be rejected by the type system. The type system can report the reason for the ill-versioned extraction by using the version information recorded in the context.

$$\textbf{let}\ [f] = \text{get_n_monitors}\ \textbf{in}\ \textbf{let}\ [x] = \{l_3 = ()\}\ \textbf{in}\ [f\ x].l_3 : (\mathit{rejected})$$

– ERROR: $f$ and $x$ are expected to be available in $l_3$, but $f$ is not available in $l_3$.

The second example is a variation of Eq. 3. The following program is rejected because $\{l_1 = ()\}$ has no definition of $l_2$.

$\textbf{let}\ [f] = \text{get_n_monitors}\ \textbf{in}\ \textbf{let}\ [x] = \{l_1 = ()\}\ \textbf{in}\ [f\ x].l_2 : (\mathit{rejected})$

– ERROR: $f$ and $x$ are expected to be available in $l_2$, but $x$ is not available in $l_2$.

## 4 The Lambda VL Type System

$\lambda_{\text{VL}}$ is an extension of the coeffect calculus $\ell\mathcal{R}$PCF [7] and GrMini [21]. We defined the *version resource algebra* and extended the coeffect calculus by adding *versioned terms*.

### 4.1 Syntax of $\lambda_{\text{VL}}$

The terms and types of $\lambda_{\text{VL}}$ are as follows:

$$t ::= \underbrace{x \mid t_1\, t_2 \mid \lambda x.t}_{\lambda\text{-terms}} \mid \underbrace{n}_{\text{constructors}} \mid \underbrace{[t] \mid \textbf{let}\ [x] = t_1\ \textbf{in}\ t_2}_{\text{coeffect terms}} \mid \underbrace{\{\overline{l = t} \mid l_i\} \mid t.l \mid \langle \overline{l = t} \mid l_i \rangle}_{\text{versioned terms}} \qquad \text{(terms)}$$

$$A, B ::= \underbrace{\mathsf{Int}}_{\text{Integer}} \mid \underbrace{A \to B}_{\text{function types}} \mid \underbrace{\Box_r A}_{\text{versioned types}} \qquad \text{(types)}$$

Many of the terms in $\lambda_{\text{VL}}$ derive from linear $\lambda$-calculus. Additional terms are categorized by those for introducing and eliminating versioned values. Versioned values can be declared through promotions $[t]$ and versioned records $\{\overline{l = t} \mid l_i\}$. The type of versioned values $\Box_r A$ are indexed by a *version resource* $r$, where $r$ ranges over the elements of the version resource semiring $\mathcal{R}$ described in section 4.2. The *versioned records* $\{\overline{l = t} \mid l_i\}$ has a *default label* along with a pair of labels and version-specific definitions. In the current design, programmers can note a default label $l_i \in \{\overline{l}\}$ for the case where there are multiple versions of a calculation results. The default label is overridden in the dynamic semantics described in section 5. The term $[t]$ is a promotion of version necessity and allows $t$ to be used to track the use of version resources in a program. The term $\textbf{let}\ [x] = t_1\ \textbf{in}\ t_2$ provides an elimination for version necessity and provides version-aware let-binding. Finally, the versioned computations $\langle \overline{l = t} \mid l_i \rangle$ represent intermediate terms whose evaluation in the default version is postponed. We assume that versioned computations appear only in intermediate terms during evaluation and not in the user's code.

### 4.2 Version Resources

The $\lambda_{\text{VL}}$ type system is parameterized by the version resource semiring $\mathcal{R}$. It captures how a program depends on its context by tracking version information on the variables used in the program. Version resources $r \in \mathcal{R}$ appear in types with $\Box$-constructors and in contexts with $[*]_r$-notions to denote the sets of versions on which the programs implicitly depend.

The version resources $r$ are given by the following:

$$r ::= \bot \mid \emptyset \mid \{l_i\} \mid r_1 \cup r_2 \qquad \text{(version resources)}$$

Intuitively, an element of $\mathcal{R}$ is a set of labels such as $\{l_1\}$ and $\{l_1, l_2\}$. The language produced by this grammar is equivalent to the elements of the version resource semiring $\mathcal{R}$.

**Definition 4.1** (Version resource semiring)**.** *The version resource semiring is given by the structural semiring (semiring with pre-order) $(\mathcal{R}, \oplus, 0, \otimes, 1, \sqsubseteq)$, defined as*

$$0 = \bot \quad 1 = \emptyset \qquad \bot \sqsubseteq r \qquad \frac{r_1 \subseteq r_2}{r_1 \sqsubseteq r_2}$$

$$r_1 \oplus r_2 = \begin{cases} r_1 & r_2 = \bot \\ r_2 & r_1 = \bot \\ r_1 \cup r_2 & \textit{otherwise} \end{cases} \qquad r_1 \otimes r_2 = \begin{cases} \bot & r_1 = \bot \\ \bot & r_2 = \bot \\ r_1 \cup r_2 & \textit{otherwise} \end{cases}$$

*where $\bot$ is the smallest element of $\mathcal{R}$, and $r_1 \subseteq r_2$ is the standard subset relation over sets defined only when both $r_1$ and $r_2$ are not $\bot$.*

The fact that version resource semiring is structural semiring with pre-order [7] is proven in the Appendix B.1.

Multiplication $\otimes$ represents that if a value is used in version $l_i$, then all values in that data flow must also be available in version $l_i$; The versioned value to resource $\{l_1\}$ can be applied with both a versioned value with resource $\{l_1\}$ and $\{l_1, l_2\}$. Addition $\oplus$ represents splitting the data flow of a value in a typing context. Therefore, all values used in version $l_i$ somewhere in the data flow must be permitted to be used in version $l_i$, even if they are used elsewhere in the context of another version $l_j$; thus, $\{l_i\} \cup \{l_j\} = \{l_i, l_j\}$.

$0 = \bot$ is the smallest element indicating an irrelevant resource. Conversely, $1 = \emptyset$ explicitly indicates that the value has no version restrictions. That is, a 1-indexed versioned value can be usable as any versioned value unless it is a 0-indexed versioned value. These intuitive explanations will be detailed later in the typing rules.

### 4.3 The Type System of $\lambda_{\mathrm{VL}}$

Typing judgments of the type system are of the form $\Gamma \vdash t : A$ with typing contexts $\Gamma$. A typing context $\Gamma$ (or we sometimes note $\Delta$) is a set of typed variables defined as follows:

$$\Gamma ::= \emptyset \mid \Gamma, x : A \mid \Gamma, x : [A]_r \qquad \text{(contexts)}$$

Typing contexts are either empty $\emptyset$ or extended with a linear variable assumption $x : A$ or a *versioned assumption* $x : [A]_r$. For a versioned assumption, $x$ can behave non-linearly, with substructural behavior captured by the semiring element $r \in \mathcal{R}$,

$\lambda_{\mathrm{VL}}$ **typing rules** $\boxed{\Gamma \vdash t : A}$

$$\frac{}{\emptyset \vdash n : \mathsf{Int}}\ (\textsc{int}) \qquad \frac{}{x : A \vdash x : A}\ (\textsc{var}) \qquad \frac{\Gamma, x : A \vdash t : B}{\Gamma \vdash \lambda x.t : A \to B}\ (\textsc{abs})$$

$$\frac{\Gamma_1 \vdash t_1 : A \to B \quad \Gamma_2 \vdash t_2 : A}{\Gamma_1 + \Gamma_2 \vdash t_1\, t_2 : B}\ (\textsc{app}) \qquad \frac{\Gamma_1 \vdash t_1 : \Box_r A \quad \Gamma_2, x : [A]_r \vdash t_2 : B}{\Gamma_1 + \Gamma_2 \vdash \mathbf{let}\ [x] = t_1\ \mathbf{in}\ t_2 : B}\ (\textsc{let})$$

$$\frac{\Gamma \vdash t : A}{\Gamma, [\Delta]_0 \vdash t : A}\ (\textsc{weak}) \qquad \frac{\Gamma, x : A \vdash t : B}{\Gamma, x : [A]_1 \vdash t : B}\ (\textsc{der}) \qquad \frac{[\Gamma] \vdash t : A}{r \cdot [\Gamma] \vdash [t] : \Box_r A}\ (\textsc{pr})$$

$$\frac{\Gamma, x : [A]_r, \Gamma' \vdash t : B \quad r \sqsubseteq s}{\Gamma, x : [A]_s, \Gamma' \vdash t : B}\ (\textsc{sub}) \qquad \frac{\Gamma \vdash t : \Box_r A \quad l \in r}{\Gamma \vdash t.l : A}\ (\textsc{extr})$$

$$\frac{[\Gamma_i] \vdash t_i : A}{\bigcup_i(\{l_i\} \cdot [\Gamma_i]) \vdash \{\overline{l = t}|l_i\} : \Box_{\{\bar{l}\}} A}\ (\textsc{ver}) \qquad \frac{[\Gamma_i] \vdash t_i : A}{\bigcup_i(\{l_i\} \cdot [\Gamma_i]) \vdash \langle \overline{l = t}|l_i \rangle : A}\ (\textsc{veri})$$

**Figure 4** $\lambda_{\mathrm{VL}}$typing rules

which describes $x$'s use in a term. We denote by $|\Gamma|$ the typing context in which every assumption is a versioned assumption.

Figure 4 shows the typing rules for $\lambda_{\mathrm{VL}}$. The typing rules for $\lambda$-terms are (INT), (VAR), (APP), and (ABS). (VAR) shows that linear variables can only be typed in a single context that includes themselves. Note that the typing rules for splitting a data flow, such as (APP), include context concatenation +, which permits the splitting version resources as defined in Def. 4.2.

**Definition 4.2** (Context concatenation , & +)**.** *Two typing contexts can be concatenated by "," if they contain disjoint sets of assumptions. Furthermore, the versioned assumptions appearing in both typing contexts can be combined using the context concatenation + defined with the addition ⊕ in the version resource semiring as follows:*

$$(\Gamma, x : A) + \Gamma' = (\Gamma + \Gamma'), x : A \quad \textit{iff}\ x \notin |\Gamma'| \qquad \emptyset + \Gamma = \Gamma$$
$$\Gamma + (\Gamma', x : A) = (\Gamma + \Gamma'), x : A \quad \textit{iff}\ x \notin |\Gamma| \qquad \Gamma + \emptyset = \Gamma$$
$$(\Gamma.x : [A]_r) + (\Gamma', x : [A]_s) = (\Gamma + \Gamma'), x : [A]_{(r \oplus s)}$$

The (WEAK) rule provides weakening only for version assumptions indexed by 0. Since $0 = \bot$ is defined as an irrelevant resource in $\mathcal{R}$, this rule indicates that adding unneeded versioned assumptions to the typing context does not prevent the term from type checking, just as in linear type systems. The (DER) rule converts a linear assumption into a versioned assumption, indexed by 1. This rule indicates the intuition that the linear assumption does not have any restrictions on versions. The (PR) rule introduces a version necessity indexed by $r$ to a term and propagates the assumption into the context using the contextual multiplication $\cdot$ defined in Def. 4.3.

**Definition 4.3** (Multiplying contexts $\cdot$ by a resource)**.** *Assuming that a context contains only version assumptions, denoted* $|\Gamma|$ *in typing rules, then* $\Gamma$ *can be multiplied by a version resource* $r \in \mathcal{R}$ *by using the product* $\otimes$ *in the version resource semiring, as follows:*

$$r \cdot \emptyset \;=\; \emptyset \qquad\qquad r \cdot (\Gamma, x : [A]_s) \;=\; (r \cdot \Gamma), x : [A]_{(r \otimes s)}$$

Informally, $r \cdot \Gamma$ requires that all assumptions in $\Gamma$ to be available in that version. This property is the cornerstone of this type system, and will be illustrated in Example 4.6 with examples.

The (SUB) rule weakens the version assumption based on the order defined in the version resource semiring. For example, suppose a value is typable in a context where a variable is only available in version 1. In that case, the value should be typable as well, even if the variable is available in both versions 1 and 2. The (SUB) rule formalizes this intuition by using the preorder in the version resource semiring. This rule is detailed in Example 4.5. The (LET) rule provides a way to remove the versioned necessity assigned to a term. From the perspective of term reuse, the version necessity assigned to a term ($\square_r A$) is converted to a version assumption ($[A]_r$) and add it to the context of the body typing. Note that the contexts of the two subterms are combined by context concatenation $+$ as well as the (APP) rule.

The last two rules (VER) and (VERI) are for version records. The context of a version record is the sum of typing contexts multiplied by the version resource corresponding to each version-specific term. Summation of typing contexts is defined as follows:

**Definition 4.4** (Context summation $\bigcup$)**.** *Using the context concatenation* $+$*, summation of typing contexts is defined as follows:*

$$\bigcup_{i \in n} \Gamma_i \;=\; \Gamma_1 + \cdots + \Gamma_n$$

In the (VERI) and the (EXTR) rules, note that the type of the versioned computation and the extraction have lost their version resource. In our current design, once a versioned value is evaluated in a particular version, it becomes a common value that can be used with other versioned values. The concluding types of these two rules illustrate this feature.

To aid in understanding of the type system, we show some important facts in the following example.

**Example 4.5** (Weakening version resources)**.**

Any linear assumption in the environment can be regarded as an arbitrary versioned assumption. This fact can be obtained by a combination of the rules (VAR), (DER), and (SUB) as follows:

$$\dfrac{\dfrac{\dfrac{}{f : \mathsf{Int} \to \mathsf{Int} \vdash f : \mathsf{Int} \to \mathsf{Int}}\;(\text{VAR})}{f : [\mathsf{Int} \to \mathsf{Int}]_1 \vdash f : \mathsf{Int} \to \mathsf{Int}}\;(\text{DER}) \qquad 1 \sqsubseteq r}{f : [\mathsf{Int} \to \mathsf{Int}]_r \vdash f : (\mathsf{Int} \to \mathsf{Int})}\;(\text{SUB})$$

This formalizes the intuition that a linear assumption does not have any constraints on the its variable use with respect to their versions.

As shown above, the (SUB) rule allows the versioned resources in a context can be increased. This fact supports the intuition that a term that is typed with versioned assumptions that is only available in a particular version will still be typed with versioned assumptions that is available in more versions.

**Example 4.6** (Ensuring the existence of consistent versions).

The purpose of the type system is to ensure that all versions needed to evaluate a given program exist. The type system ensures this property through the computation of resources with consistent shared labels. The following program, a simplified version of the program described in 3.2.5, is an example of a faulty program.

$$\textbf{let}\ [f] = \{l_1 = \mathrm{id}, l_2 = \mathrm{succ} \,|\, l_1\}\ \textbf{in}\ \textbf{let}\ [y] = \{l_1 = 1 \,|\, l_1\}\ \textbf{in}\ [f\ y].l_2$$

Type-checking this program halfway yields the following derivation tree.

$$\dfrac{\vdots \quad \dfrac{\vdots \quad \dfrac{\dfrac{\text{ERROR: } \{l_2\} \cup r \text{ cannot be a subset of } \{l_1\}}{\{l_1\}\cdot(f : [\mathsf{Int} \to \mathsf{Int}]_{\{l1,l2\}}, y : [\mathsf{Int}]_{\{l1\}}) \vdash [f\ y] : \Box_{\{l_2\}\cup r} X}\ (\text{PR})}{\{l_1\}\cdot(f : [\mathsf{Int} \to \mathsf{Int}]_{\{l1,l2\}}, y : [\mathsf{Int}]_{\{l1\}}) \vdash [f\ y].l_2 : X}\ (\text{EXTR})}{f : [\mathsf{Int} \to \mathsf{Int}]_{\{l1,l2\}} \vdash \textbf{let}\ [y] = \{l_1 = 1 \,|\, l_1\}\ \textbf{in}\ [f\ y].l_2 : X}\ (\text{LET})}{\emptyset \vdash \textbf{let}\ [f] = \{l_1 = \mathrm{id}, l_2 = \mathrm{succ} \,|\, l_1\}\ \textbf{in}\ \textbf{let}\ [y] = \{l_1 = 1 \,|\, l_1\}\ \textbf{in}\ [f\ y].l_2 : X}\ (\text{LET})$$

In this derivation tree, for clarity, the largest shared version resource is specified outside the context. Now recall that (PR) requires the same version resources for the entire context as introduced in the term. The largest shared resource in the context is $\{l_1\}$, but the resource in the term must have $\{l_2\}$ as a subset. As a result, the type checker reports this discrepancy. In this way, we use the nature of $\otimes$ to guarantee that each version value has a consistent version.

# 5 Dynamic Semantics and Metatheory

## 5.1 Dynamic Semantics

We give the small-step operational semantics of $\lambda_{\mathrm{VL}}$ in figure 5. They basically follow the lazy-evaluation strategy; i.e., only functions $t$ are evaluated to values to evaluate applications $t\ t'$. The operational semantics of $\lambda_{\mathrm{VL}}$ consist of two main parts – *evaluation* and *default version overwriting*. The $\lambda_{\mathrm{VL}}$ evaluation proceeds by alternating between reduction and default version overwriting.

We define *values* and *evaluation context* of $\lambda_{\mathrm{VL}}$ as follows:

$$v \ ::= \ \lambda x.t \ | \ n \ | \ [t] \ | \ \{\overline{l = t} \,|\, l_i\} \qquad \text{(values)}$$

$$E \ ::= \ [] \ | \ E\ t \ | \ E.l \ | \ \textbf{let}\ [x] = E\ \textbf{in}\ t \qquad \text{(evaluation contexts)}$$

**Evaluation rule**

$$\frac{t \rightsquigarrow t'}{E[t] \longrightarrow E[t']}$$

**Reduction rules**

$$\frac{}{(\lambda x.t)t' \rightsquigarrow (t' \rhd x)t}\ (\text{E-ABS}) \qquad \frac{}{\textbf{let}\ [x] = v\ \textbf{in}\ t \rightsquigarrow (v \rhd [x])t}\ (\text{E-CLET})$$

$$\frac{}{[t].l \rightsquigarrow t@l}\ (\text{E-EXI}) \qquad \frac{}{\{\overline{l = t} \,|\, m\}.l_i \rightsquigarrow t_i@l_i}\ (\text{E-EX2}) \qquad \frac{}{\langle \overline{l = t} \,|\, l_i \rangle \rightsquigarrow t_i@l_i}\ (\text{E-VERI})$$

**Substitution rules**

$$\frac{}{(t \rhd x)t' = [t/x]t'}\ (\rhd_{\text{var}}) \qquad \frac{(t \rhd x)t' = t''}{([t] \rhd [x])t' = t''}\ (\rhd_{\Box})$$

$$\frac{}{(\{\overline{l = t} \,|\, l_i\} \rhd [x])t' = [\langle \overline{l = t} \,|\, l_i \rangle / x]t'}\ (\rhd_{\text{ver}})$$

**Default version overwriting rules**

$$n@l \equiv n \qquad (\lambda x.t)@l \equiv \lambda x.(t@l) \qquad (t\,u)@l \equiv (t@l)(u@l)$$

$$\textbf{let}\ [x] = t_1\ \textbf{in}\ t_2@l \equiv \textbf{let}\ [x] = (t_1@l)\ \textbf{in}\ (t_2@l)$$

$$[t]@l \equiv [t] \qquad \{\overline{l = t} \,|\, l_i\}@l' \equiv \{\overline{l = t} \,|\, l_i\} \qquad (t.l)@l' \equiv (t@l').l$$

$$\frac{l' \in \{\bar{l}\}}{\langle \overline{l = t} \,|\, l_i \rangle @l' \equiv \langle \overline{l = t} \,|\, l' \rangle} \qquad \frac{l' \notin \{\bar{l}\}}{\langle \overline{l = t} \,|\, l_i \rangle @l' \equiv \langle \overline{l = t} \,|\, l_i \rangle}$$

**Figure 5** $\lambda_{\text{VL}}$dynamic semantics

Figure 5 shows the dynamic semantics for $\lambda_{\text{VL}}$. The $\lambda_{\text{VL}}$ has five reduction rules. The (E-ABS) rule is the $\beta$-reduction rule for the lazy evaluation strategy, and (E-CLET) is a rule for contextual let-bindings. Each uses the captured $x$ to assign a value according to the following substitution rules. Note that versioned value constructors with both variable and term are removed together in the ($\rhd_{\Box}$) and ($\rhd_{\text{ver}}$) rules. A well-typed versioned value bound to a contextual-let binding will have its outer versioned value constructors removed along with the variable. A term that will eventually be substituted into $x$ loses its outermost version resource.

The next three reduction rules are for extracting versioned values. As explained above, a versioned value can only be evaluated when it is extracted. The two versioned values – promotions and versioned records – are evaluated in terms that lose their version resources along with the @-notation by extraction. The @-notation is used to override the default version; and overwriting rules scan through the terms and overwrites the default versions of all intermediate terms $\langle \overline{l = t} \,|\, l_i \rangle$ with their label $l$. Eventually, these intermediate terms are evaluated to version-specific terms by using the (E-VERI) rule.

**Example 5.1** (Evaluation process of a versioned function application).

To aid in understanding of the dynamic semantics, we show an evaluation process of versioned function application presented in the introduction.

$$\textbf{let}\ [f] = \{l_1 = \lambda x.x, l_2 = \lambda x.x + 1 \,|\, l_1\}\ \textbf{in let}\ [y] = \{l_1 = 1, l_2 = 2 \,|\, l_1\}\ \textbf{in}\ [f\ y]$$

Both version function $f$ and version variable have different definitions with labels $l_1$ and $l_2$. As mentioned in introduction, this program intuitively evaluates to $\{l_1 = 1, l_2 = 3\}$ – precisely, to a suspended versioned computation that is expected to evaluate to 1 and 3 in each version. Hereafter, we abbreviate $\lambda x.x$ as id and $\lambda x.x + 1$ as succ.

And next, since $\{l_1 = 1, l_2 = 2 \,|\, l_1\}$ is a value, the program is evaluated as follows:

$$\begin{aligned}
&\longrightarrow \underline{[\{l_1 = \mathrm{id}, l_2 = \mathrm{succ} \,|\, l_1\} \rhd [f]]}\ (\textbf{let}\ [y] = \{l_1 = 1, l_2 = 2 \,|\, l_1\}\ \textbf{in}\ [f\ y]) && (\text{E-CLET})\\
&= \underline{[\langle l_1 = \mathrm{id}, l_2 = \mathrm{succ} \,|\, l_1\rangle / f]}\ (\textbf{let}\ [y] = \{l_1 = 1, l_2 = 2 \,|\, l_1\}\ \textbf{in}\ [f\ y]) && (\rhd_{\mathrm{ver}})\\
&= (\textbf{let}\ [y] = \{l_1 = 1, l_2 = 2 \,|\, l_1\}\ \textbf{in}\ [\langle l_1 = \mathrm{id}, l_2 = \mathrm{succ} \,|\, l_1\rangle\ y]) && (\text{substitution})
\end{aligned}$$

Note that in the first two lines, the $(\rhd_{\mathrm{ver}})$ rule simultaneously removes the versioned constructors of $[x]$ and $\{l_1 = \mathrm{id}, l_2 = \mathrm{succ} \,|\, l_1\}$. The term eventually assigned to $f$ is a versioned computation that inherits the default version $l_1$.

The program is evaluated as well for $y$.

$$\longrightarrow^* [\langle l_1 = \mathrm{id}, l_2 = \mathrm{succ} \,|\, l_1\rangle\ \langle l_1 = 1, l_2 = 2 \,|\, l_1\rangle] \qquad (\text{E-CLET}, \rhd_{\mathrm{ver}}, \text{substitution})$$

The result suspended versioned computation contains the respective computations for labels $l_1$ and $l_2$, thus we can obtain result value by extraction as shown below.

$$\begin{aligned}
&\textbf{let}\ [f] = \{l_1 = \mathrm{id}, l_2 = \mathrm{succ} \,|\, l_1\}\ \textbf{in let}\ [y] = \{l_1 = 1, l_2 = 2 \,|\, l_1\}\ \textbf{in}\ [f\ y].l_1\\
\longrightarrow^*\ & [\langle l_1 = \mathrm{id}, l_2 = \mathrm{succ} \,|\, l_1\rangle\ \langle l_1 = 1, l_2 = 2 \,|\, l_1\rangle].l_1 && (\text{E-CLET}, \rhd_{\mathrm{ver}}, \text{substitution})\\
\longrightarrow\ & (\langle l_1 = \mathrm{id}, l_2 = \mathrm{succ} \,|\, l_1\rangle\ \langle l_1 = 1, l_2 = 2 \,|\, l_1\rangle)@l_1 && (\text{E-EXI})\\
\equiv\ & \langle l_1 = \mathrm{id}, l_2 = \mathrm{succ} \,|\, l_1\rangle @l_1\ \langle l_1 = 1, l_2 = 2 \,|\, l_1\rangle @l_1 && (\text{def-ver overwriting})\\
\equiv^*\ & 1
\end{aligned}$$

We can perform the same kind of extraction for $l_2$ and obtain 3.

### 5.2 Metatheory

We give a precise formalization to some properties of $\lambda_{\mathrm{VL}}$. As with other coeffect calculi, there are two variants of the substitution lemma, one through linear assumptions and the other through versioned assumptions. The proofs are given by structural induction on the typing derivation and are somewhat tricky. We have to carefully manage how version resources are divided in the typing context; thus, we would like to adopt a generalized form of the versioned substitution lemma.

**Lemma 5.2** (Well-typed linear substitution). *Let* $\Delta \vdash t' : A$ *and* $\Gamma, x : A, \Gamma' \vdash t : B$. *Then,* $\Gamma + \Delta + \Gamma' \vdash [t'/x]t : B$

**Subsumption rules**

$$\frac{}{A <: A} \qquad \frac{A <: B \quad r' \sqsubseteq r}{\Box_r A <: \Box_{r'} B} \qquad \frac{A' <: A \quad B <: B'}{A \to B <: A' \to B'}$$

$$\frac{A <: B \quad r' \sqsubseteq r}{[A]_r <: [B]_{r'}} \qquad \frac{}{\Gamma \sqsubseteq \Gamma} \qquad \frac{\Gamma \sqsubseteq \Delta \quad A <: B}{\Gamma, x : B \sqsubseteq \Delta, x : A}$$

**Figure 6** Subsumption rules for typing context

**Lemma 5.3** (Well-typed versioned substitution). *Let* $[\Delta] \vdash t' : A$ *and* $\Gamma, x : [A]_r, \Gamma' \vdash t : B$. *Then,* $\Gamma + \bigcup_i (r_i \cdot [\Delta_i]) + \Gamma' \vdash [t'/x]t : B$ *where* $\Sigma_i r_i = r$ *and* $\bigcup_i [\Delta_i] = \Delta$

In most cases, this generalization is unaffected, i.e., $\bigcup_i (r_i \cdot [\Delta_i]) = r \cdot \Delta$, but in the case where the last derivation of $t$ is derived from (VER)-rule, this difference becomes essential.

By using the two substitution lemmas, we give proof of $\lambda_{\text{VL}}$ type safety.

**Theorem 5.4** ($\lambda_{\text{VL}}$ progress). *Let* $\Gamma \vdash t : A$. *Then, (i)* $t$ *is a value or (ii)* $\exists t'.t \longrightarrow t'$

**Theorem 5.5** ($\lambda_{\text{VL}}$ preservation). *Let* $\Gamma \vdash t : A$ *and* $t \longrightarrow t'$. *Then,* $\exists \Gamma'.\Gamma' \vdash t' : A \wedge \Gamma' \sqsubseteq \Gamma$

The conclusions of the above two theorems allow for an order between typing contexts described in figure 6. This ordering relation $\sqsubseteq$ intuitively implies that for any assumption in $\Gamma'$, the version resource is less than that in $\Gamma$ such as $x : [A]_{\{l1\}}, y : [A]_{\{l2\}} \sqsubseteq x : [A]_{\{l1,l2\}}, y : [A]_{\{l1,l2\}}$.

# 6 Related Work

## 6.1 Software Product Lines

Software Product Lines (SPLs) [22, 25] are methods for creating a collection of similar software from a shared set of programs. Since program updates can be considered as a kind of program extension, some programming techniques in SPLs [6, 26, 30] are closely related to this work.

### 6.1.1 Delta-Oriented Programming

Delta-oriented programming (DOP) [29, 30, 31] provides a mechanism called *delta-modules* for modularizing program modifications. The delta-module language allows not only the addition and overriding of classes and methods, but also their removal. Each delta module contains the application conditions for modifications, and delta modules can be combined to form complex constraints on the features of a product.

We think that the DOP approach is complementary to our research. For implementation of version analysis by type checking in the future, it is essential to have a packaging system with expression-level dependency information. Given that the evolution of packages today is basically linear over versions, it should be possible to develop a package system modularized by program diffs. Patrick Lam *et al.* [18] point out that the lack of tool support for package changes requires developers to pay a great deal of attention to compatibility, and discuss the implications of calculating compatibility information in the context of program analysis. For implementation of version analysis by type checking in the future, it is essential to have a packaging system with expression-level dependency information. We believe that such tools will lead to the development of a novel package system with more detailed information about compatibility.

#### 6.1.2 Variational Programming

Variational Programming [15, 34, 35] is a language paradigm with syntactic support for data variation. For example, the Variational Programming Calculus [8] (VPC) represents differences as binary choices called *dimensions*, such as `A<1,true>`, which can be passed as function arguments. In addition, we can extract and manipulate choices. For example, a program that applies the identity function to either 1 or `true` and extracts left one can be written as `sel A.L (λx.x A<1, true>)`.

At first glance, it seems that the main part of the $\lambda_{\text{VL}}$ can be encoded by VPC, but this is impossible. The key difference is semantics: VPC doesn't have a mechanism to deal with computations that lack definitions like versioned records in $\lambda_{\text{VL}}$. In $\lambda_{\text{VL}}$, functions and arguments with different dimensions are applied according to a smaller dimension: the calculation is allowed by the type system only if they have common versions, i.e. $\textbf{let}\ [x] = \{v1 = 1, v2 = 2\}\ \textbf{in let}\ [y] = \{v1 = 1, v2 = 2, v3 = 3\}\ \textbf{in}\ [x+y]$ is well-typed and interpreted as $\{v1 = 1+1, v2 = 2+2\}$ defined only in versions $v1$ and $v2$. In contrast, in VPC, such computations are interpreted according to a larger dimension: functions/values with smaller dimensions are interpreted in a distributed manner, i.e. `A1<1, 2> + A1<1, A2<2, 3> >` is interpreted as `A1<1 + 1, A2<2 + 2, 3 + 2> >` that have three variations. Since our primary interest is in disallowing programs to run in versions for which no definition exists, the semantics of VPC don't meet our motivations.

### 6.2 Coeffect Calculus

Coeffect calculus simultaneously arose from several contexts in the literature [7, 16, 23] since 2010s. The common denominator of these formalizations is that they annotate the assumptions in the typing context with usage information derived from a semiring. Recent studies use coeffect calculus to track bounded reuse [7, 23], deconstructor use [24], security levels [21], and scheduling constraints in hardware synthesis [16].

Granule [21] is a fully-fledged functional language focused on coeffect calculus, and its core Gr demonstrates a good combination between coeffect calculus and standard language features, *i.e.,* data types, pattern matching, and recursion. Since the core subset of Granule, GrMini, has almost the same structure as $\lambda_{\text{VL}}$, we expect that most of the language extensions from GrMini to Gr can be applied to $\lambda_{\text{VL}}$. On the other hand, the difference is that the only thing that affects resources in these languages is the availability of each functions, whereas $\lambda_{\text{VL}}$ provides a means to manipulate resources such as version records and extractions.

# 7 Future Work and Conclusion

## 7.1 Towards a Per-expression Dependency Analysis

The current $\lambda_{\mathrm{VL}}$ type system cannot express the range of compatibility that cargo and npm do in packages. For example, even if the versions 1.0.0 and 1.0.1 of package $P$ are compatible and we apply the value of type $\square_{1.0.1}P.T$ to the function that expects an argument of type $\square_{1.0.0}P.T$, this function application is rejected. This is because there is no relationship between types annotated with version resource 1.0.0 and 1.0.1 as the current type system focuses on preventing computation on versions for which a definition does not exist. Considering that many updates change only a small part of the package code and remain backward compatible for the most part, the current type system is too strict.

The next step of our research is to track the range of compatibility in the type system. Incorporating semantic versioning into types is a promising idea. Semantic versioning can also be seen as a compatibility contract from the package provider to the package users. For example, a 1.0.1 package is guaranteed to be backward compatible with 1.0.0. From the point of view of Liskov's substitution principle [19], emulating the semantic versioning strategy at the expression level, it is possible to regard $\square_{1.0.1}A$ as a subtype of $\square_{1.0.0}A$. Such a type system paves the way for type-safe casts between objects derived from different-version packages.

## 7.2 Implementation and Further Language Extensions

One future challenge is to develop an efficient implementation. Since our semantics are based on a lazy evaluation strategy, it will not be easy to implement them efficiently with the current system.

One possible approach is to implement a mechanism for version analysis in the existing coeffect language. The core calculus of the Granule language demonstrates the simultaneous use of various computational resources in the same type system scheme. Furthermore, recent work [9, 10] seeks to integrate the coeffects calculus with dependent type systems, which allow user-defined resource algebras to provide internally extensible systems in Granule [21]. These enable version analysis to be incorporated into the coeffect language in a more sophisticated form.

Another possible approach is to implement version analysis as a preprocess in the compilation of an existing language. This approach would require defining a meta-language with a type system to be versioned and would need to consider interactions with more advanced language features (class inheritance, higher-order polymorphism, etc.). We expect that if we can add version information to each value and function in a package interface with extended syntax, we can aggregate more detailed compatibility information by type checking. The advantage of this approach is that once type checking has identified the single version required for each program, we can reuse the existing runtime. We are currently trying to apply our findings from $\lambda_{\mathrm{VL}}$ to the ML-like language.

### 7.3 Conclusion

Even though dependency hell has been considered a problem for many years, it is still difficult for most programmers to solve. Most recent build tools use compatibility maintenance strategies like semantic versioning, but name mangling may burden the development community in programming languages with sophisticated type systems.

Our research aims to enable programmers to more freely combine and control programs of different versions in a single code. This research brings versions, which used to be simply identifiers of packages, into a programming language, and provides a new perspective on handling multiple versions of programs. As a first step toward our goal, we discussed the type safety of programs with multiple versions in $\lambda_{\text{VL}}$. We hope that this research will stimulate a discussion in the research community on compatibility in the context of program analysis.

**Acknowledgements** We thank the lab members at the Tokyo Institute of Technology for valuable discussions that helped shape the discussion in the paper. Especially, Youyou Cong provided feedback on earlier versions of the paper. We also thank Patrick Rein and Marcel Taeumel, Jens Lincke, and Robert Hirschfeld for discussions about the work presented in the paper. Ken Wakita, Kazuyuki Shudo, and Yasuhiko Minamide provided valuable comments in earlier steps of the work. This work was supported by JSPS KAKENHI grant numbers JP18H03219 and 20K21790.

## A Definitions

**$\lambda_{\text{VL}}$ syntax**

$$
\begin{aligned}
t &::= x \mid t_1\, t_2 \mid \lambda x.t \mid n \mid [t] \mid \textbf{let}\,[x] = t_1\ \textbf{in}\ t_2 \\
&\quad \mid \{\overline{l = t} \,|\, l_i\} \mid t.l \mid \langle \overline{l = t} \,|\, l_i \rangle && \text{(terms)} \\
v &::= \lambda x.t \mid n \mid [t] \mid \{\overline{l = t} \,|\, l_i\} && \text{(values)} \\
A, B &::= \mathsf{Int} \mid A \to B \mid \Box_r A && \text{(types)} \\
\Gamma, \Delta &::= \emptyset \mid \Gamma, x : A \mid \Gamma, x : [A]_r && \text{(contexts)} \\
r &::= \bot \mid \emptyset \mid \{l_i\} \mid r_1 \cup r_2 && \text{(version resources)} \\
E &::= [\,] \mid E\ t \mid E.l \mid \textbf{let}\,[x] = E\ \textbf{in}\ t && \text{(evaluation contexts)}
\end{aligned}
$$

**Definition A.1** (Version resource semiring)**.** *The resource algebra is given by the structural semiring (semiring with pre-order) $(\mathcal{R}, \oplus, 0, \otimes, 1, \sqsubseteq)$, defined as*

$$
0 = \bot \quad 1 = \emptyset \quad \bot \sqsubseteq r \quad \frac{r_1 \subseteq r_2}{r_1 \sqsubseteq r_2}
$$

$$
r_1 \oplus r_2 = \begin{cases} r_1 & r_2 = \bot \\ r_2 & r_1 = \bot \\ r_1 \cup r_2 & \textit{otherwise} \end{cases} \qquad r_1 \otimes r_2 = \begin{cases} \bot & r_1 = \bot \\ \bot & r_2 = \bot \\ r_1 \cup r_2 & \textit{otherwise} \end{cases}
$$

*where $\bot$ are the smallest element of $\mathcal{R}$, and $r_1 \subseteq r_2$ is the standard subset relation over sets defined only when both $r_1$ and $r_2$ are not $\bot$.*

**Definition A.2** (Context concatenation , & +)**.** *Two typing contexts can be concatenated by "," if they contain disjoint sets of linear assumptions. Furthermore, the versioned assumptions appearing in both typing contexts can be combined using the addition $\oplus$ defined in the version resource semiring. We define the context concatenation $+$ as follows:*

$$
\begin{aligned}
(\Gamma, x : A) + \Gamma' &= (\Gamma + \Gamma'), x : A \quad \textit{iff } x \notin |\Gamma'| & \emptyset + \Gamma = \Gamma \\
\Gamma + (\Gamma', x : A) &= (\Gamma + \Gamma'), x : A \quad \textit{iff } x \notin |\Gamma| & \emptyset + \Gamma = \Gamma \\
(\Gamma.x : [A]_r) + (\Gamma', x : [A]_s) &= (\Gamma + \Gamma'), x : [A]_{(r \oplus s)}
\end{aligned}
$$

**Definition A.3** (Multiplying contexts $\cdot$ by a resource)**.** *Assuming that a context contains only version assumptions, denoted $|\Gamma|$ in typing rules, then $\Gamma$ can be multiplied by a version resource $r \in \mathcal{R}$ by using the product $\oplus$ in the version resource semiring, as follows:*

$$
r \cdot \emptyset = \emptyset \qquad r \cdot (\Gamma,\, x : [A]_s) = (r \cdot \Gamma),\, x : [A]_{(r \otimes s)}
$$

**Definition A.4** (Context summation $\biguplus$)**.** *Using the context concatenation $+$, summation of typing contexts is defined as follows:*

$$
\biguplus_{i \in n} \Gamma_i = \Gamma_1 + \cdots + \Gamma_n
$$

## B Proofs

**Lemma B.1** (Version resource). *Version resource semiring $(\mathscr{R}, \oplus, \bot, \otimes, \emptyset, \sqsubseteq)$ is a structural semiring.*

*Proof.* Version resource semiring $(\mathscr{R}, \oplus, \bot, \otimes, \emptyset, \sqsubseteq)$ induces a semilattice with $\oplus$ (join).

- $(\mathscr{R}, \oplus, \bot, \otimes, \emptyset)$ is a semiring, that is:
  - $(\mathscr{R}, \oplus, \bot)$ is a commutative monoid, i.e., for all $p, q, r \in \mathscr{R}$
    - (Associativity) $(p \oplus q) \oplus r = p \oplus (q \oplus r)$ holds since $\oplus$ is defined in associative manner with $\bot$.
    - (Commutativity) $p \oplus q = q \oplus p$ holds since $\oplus$ is defined in commutative manner with $\bot$.
    - (Identity element) $\bot \oplus p = p \oplus \bot = p$
  - $(\mathscr{R}, \otimes, \emptyset)$ is a monoid, i.e., for all $p, q, r \in \mathscr{R}$
    - (Associativity) $(p \otimes q) \otimes r = p \otimes (q \otimes r)$ holds since $\oplus$ is defined in associative manner with $\bot$.
    - (Identity element) $\emptyset \otimes p = p \otimes \emptyset = p$
      - if $p = \bot$ then $\emptyset \otimes \bot = \bot \otimes \emptyset = \bot$
      - otherwise if $p \neq \bot$ then $\emptyset \otimes p = \emptyset \cup p = p$ and $p \otimes \emptyset = p \cup \emptyset = p$
  - multiplication $\otimes$ distributes over addition $\oplus$, i.e., for all $p, q, r \in \mathscr{R}, r \otimes (p \oplus q) = (r \otimes p) \oplus (r \otimes q)$ and $(p \oplus q) \otimes r = (p \otimes r) \oplus (q \otimes r)$
    - if $r = \bot$ then $r \otimes (p \oplus q) = \bot$ and $(r \otimes p) \oplus (r \otimes q) = \bot \oplus \bot = \bot$.
    - otherwise if $r \neq \bot$ and $p = \bot$ and $q \neq \bot$ then $r \otimes (p \oplus q) = r \otimes q = r \cup q = (r \cup r) \cup q = r \cup (r \cup q) = (r \oplus p) \cup (r \cup q) = (r \otimes p) \oplus (r \otimes q)$
    - otherwise if $r \neq \bot$ and $p = \bot$ and $q = \bot$ then $r \otimes (p \oplus q) = r \otimes \bot = \bot$ and $(r \otimes p) \oplus (r \otimes q) = \bot \oplus \bot = \bot$.
    - otherwise if $r \neq \bot$ and $p \neq \bot$ and $q \neq \bot$ then $r \otimes (p \oplus q) = r \cup (p \cup q) = (r \cup p) \cup (r \cup q) = (r \otimes p) \oplus (r \otimes q)$

    The other cases are symmetrical cases.
  - $\bot$ is absorbing for multiplication: $p \otimes \bot = \bot \otimes p = \bot$ for all $p \in \mathscr{R}$
- $(\mathscr{R}, \sqsubseteq)$ is a bounded semilattice, that is
  - $\sqsubseteq$ is a partial order on $\mathscr{R}$ such that the least upper bound of every two elements $p, q \in \mathscr{R}$ exists and is denoted by $p \oplus q$.
  - there is a least element; for all $r \in \mathscr{R}$, $\bot \sqsubseteq r$.
- (Motonicity of $\oplus$) $p \sqsubseteq q$ implies $p \oplus r \sqsubseteq q \oplus r$ for all $p, q, r \in \mathscr{R}$
  - if $r = \bot$ then $p \oplus r \sqsubseteq q \oplus r \iff \bot \subseteq \bot$, so this case is trivial.
  - otherwise if $r \neq \bot, p = q = \bot$ then $p \oplus r \sqsubseteq q \oplus r \iff \bot \subseteq \bot$, so this case is trivial.
  - otherwise if $r \neq \bot, p = \bot, q \neq \bot$ then $p \oplus r \sqsubseteq q \oplus r \iff \bot \subseteq q \cup r$, so this case is trivial.
  - otherwise if $r \neq \bot, p \neq \bot, q \neq \bot$ then $p \oplus r \sqsubseteq q \oplus r \iff p \cup r \subseteq q \cup r$, and $p \subseteq q$ implies $p \cup r \subseteq q \cup r$.
- (Motonicity of $\otimes$) $p \sqsubseteq q$ implies $p \otimes r \sqsubseteq q \otimes r$ for all $p, q, r \in \mathscr{R}$

- if $r = \bot$ then $p \otimes r \sqsubseteq q \otimes r \Leftrightarrow p \subseteq q$, so this case is trivial.
- otherwise if $r \neq \bot, p = q = \bot$ then $p \otimes r \sqsubseteq q \otimes r \Leftrightarrow r \subseteq r$, so this case is trivial.
- otherwise if $r \neq \bot, p = \bot, q \neq \bot$ then $p \otimes r \sqsubseteq q \otimes r \Leftrightarrow r \subseteq q \cup r$, and $r \subseteq q \cup r$ holds in standard subset relation.
- otherwise if $r \neq \bot, p \neq \bot, q \neq \bot$ then $p \otimes r \sqsubseteq q \otimes r \Leftrightarrow p \cup r \subseteq q \cup r$, and $p \subseteq q$ implies $p \cup r \subseteq q \cup r$.

$\square$

**Lemma B.2** (Well-typed linear substitution). *Let* $\Delta \vdash t' : A$ *and* $\Gamma, x : A, \Gamma' \vdash t : B$. *Then,* $\Gamma + \Delta + \Gamma' \vdash [t'/x]t : B$

*Proof.* By induction on the derivation of $\Gamma, x : A, \Gamma' \vdash t : B$. $\square$

**Lemma B.3** (Well-typed versioned substitution). *Let* $[\Delta] \vdash t' : A$ *and* $\Gamma, x : [A]_r, \Gamma' \vdash t : B$. *Then,* $\Gamma + \bigcup_i (r_i \cdot [\Delta_i]) + \Gamma' \vdash [t'/x]t : B$ *where* $\Sigma_i r_i = r$ *and* $\bigcup_i [\Delta_i] = \Delta$

*Proof.* By induction on the derivation of $\Gamma, x : [A]_r, \Gamma' \vdash t : B$. $\square$

**Lemma B.4** (Default version overwriting type safety). *Let* $[\Gamma] \vdash t' : A$. *Then,* $\exists t'. t@l \equiv t' \wedge \{l\} \cdot [\Gamma] \vdash t' : A$

*Proof.* By induction on the derivation of $[\Gamma] \vdash t' : A$. $\square$

**Theorem B.5** (*VL* type safety). *Let* $\Gamma \vdash t : A$. *Then, (i)* $t$ *is a value or (ii)* $\exists t', \Gamma'. t \rightsquigarrow t' \wedge \Gamma' \vdash t' : A' \wedge \Gamma' \sqsubseteq \Gamma$

*Proof.* By induction on the derivation of $\Gamma \vdash t : A$. $\square$

## About the authors

**Yudai Tanabe** is a Ph.D. student in Tokyo Institute of Technology. His research interest is types and programming languages. Contact him at yudaitnb@prg.is.titech.ac.jp.

**Luthfan Anshar Lubis** is a first-year PhD student in Tokyo Institute of Technology working on programming language. Contact him at luthfanlubis@prg.is.titech.ac.jp.

**Tomoyuki Aotani** is a consultant at Mamezou Co., Ltd. His research interests include modularity, static and dynamic semantics, program analysis, program optimization, and automated program repair. Contact him at tomoyuki-aotani@mamezou.com.

**Hidehiko Masuhara** is a Professor of Mathematical and Computing Science at Tokyo Institute of Technology. His research interest is programming languages, especially on aspect- and context oriented programming, partial evaluation, computational reflection, meta-level architectures, parallel/concurrent computing, and programming environments. Contact him at masuhara@acm.org.